\documentclass[journal]{IEEEtran}
\usepackage{amsmath,bm}

\usepackage{cite}

\ifCLASSINFOpdf
  \usepackage[pdftex]{graphicx}
\else
\fi
\usepackage{amsmath}
\usepackage{amssymb}
\usepackage{array}

\usepackage[caption=false,font=footnotesize]{subfig}

\begin{document}
%
\title{Adaptive Channel-Matched Detection for C-Band 64-Gbit/s Optical OOK System over 100-km Dispersion-Uncompensated Link}
%
%
%

\author{Haide~Wang,
        Ji~Zhou,
        Dong~Guo,
        Yuanhua~Feng,
        Weiping~Liu,
        Changyuan~Yu,
        and~Zhaohui~Li

\thanks{Manuscript received; revised. This work was supported in part by National Key R\&D Program of China (2018YFB1800902); Leading Talents Program of Guangdong Province (00201502); Natural Science Foundation of Guangdong Province (2019A1515011059); Science and Technology Planning Project of Guangdong Province (2018B010114002); Local Innovation and Research Teams Project of Guangdong Pearl River Talents Program (2017BT01X121); National Natural Science Foundation of China (61525502, 61705088); Fundamental Research Funds for the Central Universities (21619309); Open Fund of IPOC (BUPT) (IPOC2019A001). (Corresponding authors: Ji Zhou and Zhaohui Li.)}

\thanks{H. Wang, J. Zhou, Y. Feng and W. Liu are with Department of Electronic Engineering, College of Information Science and Technology, Jinan University, Guangzhou 510632, China (e-mail: 1834041007@stu2018.jnu.edu.cn; zhouji@jnu.edu.cn; favinfeng@163.com; wpl@jnu.edu.cn).}

\thanks{D. Guo is with School of Information and Electronics, Beijing Institute of Technology, Beijing 100081, China (e-mail: 7520190141@bit.edu.cn).}

\thanks{C. Yu is with the Department of Electronic and Information Engineering, The Hong Kong Polytechnic University, Hong Kong (e-mail: changyuan.yu@polyu.edu.hk).}

\thanks{Z. Li is with the State Key Laboratory of Optoelectronic Materials and Technologies, School of Electronics and Information Technology, Sun Yat-sen	University, Guangzhou 510275, China and Southern Marine Science and Engineering Guangdong Laboratory (Zhuhai), Zhuhai, China (e-mail: lzhh88@mail.sysu.edu.cn).}}
%
%

\markboth{}%
{Shell \MakeLowercase{\textit{et al.}}: Bare Demo of IEEEtran.cls for IEEE Journals}
%



\maketitle

\begin{abstract}
In this paper, we propose adaptive channel-matched detection (ACMD) for C-band 64-Gbit/s intensity-modulation and direct-detection (IM/DD) optical on-off keying (OOK) system over a 100-km dispersion-uncompensated link. The proposed ACMD can adaptively compensate most of the link distortions based on channel and noise characteristics, which includes a polynomial nonlinear equalizer (PNLE), a decision feedback equalizer (DFE) and maximum likelihood sequence estimation (MLSE). Based on the channel characteristics, PNLE eliminates the linear and nonlinear distortions, while the followed DFE compensates the spectral nulls caused by chromatic dispersion. Finally, based on the noise characteristics, a post filter can whiten the noise for implementing optimal signal detection using MLSE. To the best of our knowledge, we present a record C-band 64-Gbit/s IM/DD optical OOK system over a 100 km dispersion-uncompensated link achieving 7\% hard-decision forward error correction limit using only the proposed ACMD at the receiver side. In conclusion, ACMD-based C-band 64-Gbit/s optical OOK system shows great potential for future optical interconnects.
\end{abstract}


%
\IEEEpeerreviewmaketitle

\section{Introduction}
%
%

\IEEEPARstart{D}{riven} by the continuous growth in the data demand of Internet applications, such as high definition television, online gaming, and cloud computing, etc., the transmission capacity of data center interconnects (DCI) has experienced a dramatic growth \cite{zhong2018digital, cheng2018recent, hu2019im/dd}. The optical interconnects for DCI ranges from several hundred meters to a hundred kilometers. Owing to the characteristics of low cost, low power consumption and small footprint, intensity-modulation and direct-detection (IM/DD) optical interconnects have received many attentions as the mainstream technology solutions for DCI within 100 km \cite{yan201480, zhang2015c-band, zhang2017eml-based}. In C-band IM/DD optical interconnects, chromatic dispersion (CD)-induced nulls on the signal spectrum is the main obstacle to limit the achievable capacity-distance product. For mitigating the CD-caused nulls, O-band IM/DD optical interconnects with zero dispersion have been widely studied \cite{Gao20172,miao2017experimental}. However, O-band IM/DD optical interconnects has a  transmission loss of 0.32 dB/km, which limits the received optical signal-to-noise ratio (OSNR), especially for the beyond 40-km transmission distance. Therefore, for achieving 100-km transmission distance, C-band IM/DD optical interconnects are preferred and the compensation for CD-induced nulls is crucial.

\begin{figure*}
	\centering
	\includegraphics[width = 6.8 IN]{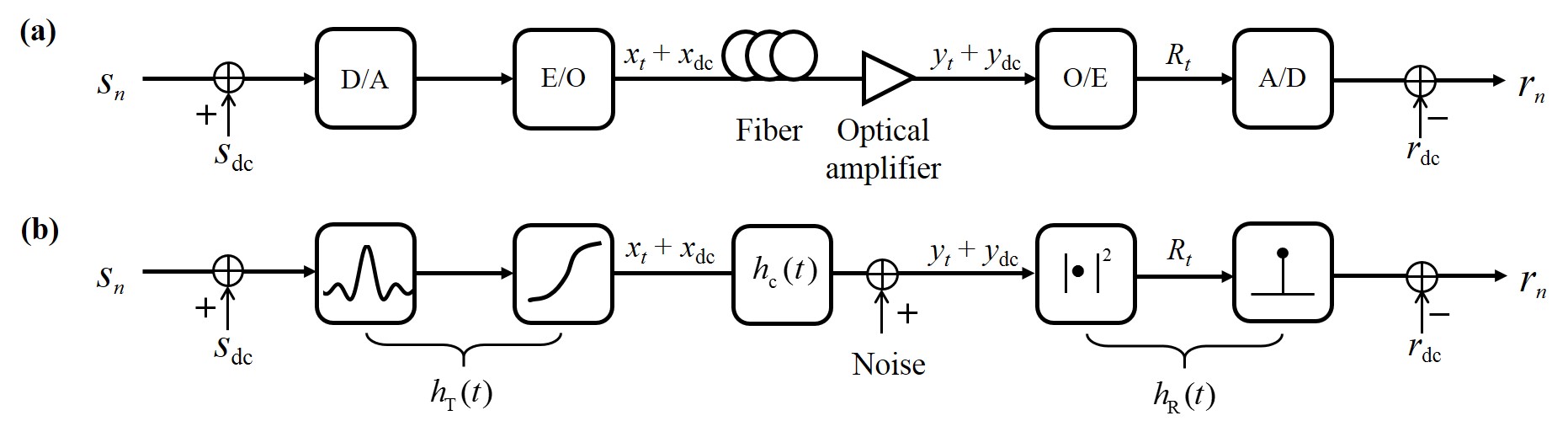}
	\caption{(a) The block diagram of a general IM/DD optical fiber communication system and (b) its mathematical description in the equivalent baseband.}
	\label{Block diagram}
\end{figure*}

The CD-induced distortions can be eliminated by using dispersion compensation module \cite{eiselt2016evaluation}, single-sideband modulation \cite{wan201764-gb/s}, CD pre-compensation \cite{zhang2017single} or Kramers-Kronig receiver \cite{mecozzi2016kramers}. However, either extra devices or complicated transmitter configurations are required in these CD compensation techniques, which increases cost and reduce flexibility of optical interconnects. With the advent of low-cost and high-speed digital signal processing (DSP), digital CD compensation is becoming cost-effective solution to flexibly mitigate CD-induced distortions for optical interconnects. The popular digital CD compensation includes decision feedback equalizer (DFE) \cite{tang2019c}, Tomlinson-Harashima pre-coding \cite{hu201850, rath2017tomlinson} and maximum likelihood sequence estimation (MLSE)\cite{zhou2020c}. However, the existing digital CD compensation techniques only consider a part of channel or noise characteristics, which cannot yet achieve the performance of optimal detection for the received signal with CD-caused distortions.

In this paper, we propose adaptive channel-matched detection (ACMD) for experimentally implementing C-band 64-Gbit/s optical on-off keying (OOK) system over a 100-km dispersion-uncompensated link. The main contributions of this paper are as follows:
\begin{itemize}
\item The ACMD is first proposed to compensate most of the link distortions by fully considering the main channel and noise characteristics. In ACMD, polynomial nonlinear equalizer (PNLE) eliminates the linear and nonlinear channel distortions, while the followed DFE compensates the spectral nulls caused by CD. Based on the noise characteristics, a post filter (PF) whitens the colored noise before realizing optimal detection using MLSE.
\item To the best of our knowledge, we present a record C-band 64-Gbit/s IM/DD optical OOK system over a 100 km dispersion-uncompensated link achieving 7\% hard-decision forward error correction (HD-FEC) limit using only receiver-side ACMD.
\end{itemize}

The remainder of the paper is organized as follows. In Section \ref{Principle}, the distortions model of a general IM/DD optical fiber communication system and the principle of the proposed ACMD are given. In Section \ref{ESetups},  we present the experimental setups of C-band 64-Gbit/s IM/DD optical OOK transmission over dispersion-uncompensated links using ACMD. The experimental results and discussions are demonstrated in Section \ref{ER}. Finally, the paper is concluded in Section \ref{Conclusion}.

%


\section{Principle}
\label{Principle}
\subsection{Distortions Model Of IM/DD Systems}

Figure \ref{Block diagram}(a) shows the block diagram of a general IM/DD optical fiber communication system. As shown in Fig. \ref{Block diagram}(a), after adding a direct current (DC) value $s_{\text{dc}}$ to the bipolar transmitted sequence $s_n$, the unipolar modulation can be achieved. Then the unipolar transmitted digital signal is reconstructed to analog signal by digital-to-analog (D/A) conversion. After the electrical-to-optical (E/O) conversion, the optical transmitted bipolar signal $x_t$ and the DC component $x_{\text{dc}}$ are uploaded into the fiber and amplified by an optical amplifier. At the receiver, after the optical-to-electrical (O/E) conversion, optical received bipolar signal $y_t$ and the DC component $y_{\text{dc}}$ are converted to the real-valued analog signal $R_t$. Then the digital signal is obtained by analog-to-digital (A/D) conversion. Finally, after the subtraction of a DC value $r_{\text{dc}}$, the bipolar received sequence $r_n$ is send to the following DSP and hard decision.

The mathematical description of the general IM/DD optical fiber communication system in the equivalent baseband is shown in Fig. \ref{Block diagram}(b). The transmitter and receiver are described by the impulse responses $h_{\text{T}}(t)$ and $h_{\text{R}}(t)$ of low-pass models, respectively. The fiber and optical amplifier are described by the impulse response $h_{\text{C}}(t)$. The major impairments limiting the performance of IM/DD systems include the following:
\subsubsection{Linear and nonlinear distortions}
Due to the bandwidth limitation of the devices, the signal suffers from linear inter-symbol interference (ISI). Since the optical channel is nonlinear, the high launch power will result in strong nonlinear effect. The E/O modulator will introduce nonlinear distortion to the signal. In addition, regardless of the noise, the real-valued signal after the absolute square law detection of the O/E device (e.g., photo-diode) is
\begin{equation}
R_t \propto y_{\text{dc}}^{2} + 2y_{\text{dc}}x_t \mathcal{F}^{-1}\left\{\operatorname{real}\left\{H_{\text{C}}(\omega)\right\}\right\} + |x_t \otimes h_{\text{C}}(t)|^{2}
\label{Eq3}
\end{equation}
where $\otimes$ denotes convolution, $\mathcal{F}^{-1}\{\cdot\}$ is the inverse fast Fourier transform and $\operatorname{real}\{\cdot\}$ is the real component of a complex number. $x_{\text{dc}}^{2}$ is a DC component. The third term of right side of Eq.(\ref{Eq3}) is signal-signal beat interference (SSBI). Although there is little influence on the OOK modulation, the SSBI introduces nonlinear distortions to the advanced modulation formats, such as multilevel pulse amplitude modulation (M-PAM), discrete multitone modulation, carrierless amplitude and phase modulation.

\subsubsection{CD}
CD imposes a severe fading effect on the power of received signal, which is the major transmission impairment. For simplicity, channel characteristics $H_{\text{C}}(\omega)$ can be roughly characterized by optical fiber. The low-pass equivalent model for the standard single mode fiber (SSMF) frequency response can be expressed as \cite{elrefaie1991chromatic}
\begin{equation}
H_{\text{SSMF}}(\omega) = e^{j\beta_2L\omega^2/2}
\end{equation}
where $\beta_2$ denotes the group velocity dispersion, $L$ is the length of SSMF and $\omega$ is the angular frequency parameter. Therefore, the second term of right side of Eq.(\ref{Eq3}) can be roughly express as $ 2 y_{\text{dc}} x_{t} \mathcal{F}^{-1}\{\cos(\beta_2L\omega^2/2)\}$, which refers to power-fading effect caused by CD. The spectrum of the received signal has spectral nulls due to the CD, which will greatly degrade the system performance.

\subsubsection{Noise}
Fiber loss has been compensated greatly because of the development of the optical amplifiers. Erbium doped fiber amplifier (EDFA) is usually adopted to amplify the signal but simultaneously introduces amplified spontaneous emission (ASE) noise. In the optical fiber channel, ASE noise in the optical amplifier is the major source of noise. Additionally, the noise introduced by the devices, including thermal noise, shot noise and quantizing noise, is amplified by the optical amplifier in optical domain at the same time. The noise that greatly increases the probability of error in symbol decision cannot be totally eliminated by any algorithm. Noise is the hardest impairments to be settled.

\subsection{Proposed ACMD For IM/DD Systems}
\label{ACMDPrinciple}
When it successfully integrates the unique capabilities of mathematics with the insights and prior information gained from the underlying physics of the problem, signal processing is at its best \cite{haykin2001signal}. Based on the above analysis of the distortions model of IM/DD systems, ACMD is proposed to compensate the signal by matching the channel adaptively for signal detection, which includes the following three parts:

\subsubsection{PNLE}
Feedforward equalizer (FFE) has been successfully employed to mitigate the linear distortions. The output of the FFE is expressed as
\begin{equation}
p_n= \sum_{k=0}^{K-1} h_{k} \cdot r_{n-k}
\end{equation}
where $h_k$ is the tap coefficients and $K$ is the tap number.
However, in the IM/DD systems, except for linear distortions, nonlinear distortions exist. For eliminating the linear and nonlinear distortions of the signal, Volterra nonlinear equalizer (VNLE) is usually adopted in IM/DD systems and coherent systems as well \cite{xia2005performance, chen2016224, gao2016112}. In optical IM/DD systems, the $3^{\text{rd}}$-order VNLE is sufficient to compensate electrical and optical linear and nonlinear distortions. The output of the $3^{\text{rd}}$-order VNLE can be expressed as
\begin{equation}
\begin{aligned}
p_{n} &=\sum_{k=0}^{K_1 - 1} h_{k}^{(1)}\cdot r_{n-k}+\sum_{k=0}^{K_2 - 1} \sum_{l=0}^{k} h_{k, l}^{(2)}\cdot r_{n-k} r_{n-l} \\
&+\sum_{k=0}^{K_3-1} \sum_{l=0}^{k} \sum_{m=0}^{l} h_{k, l, m}^{(3)}\cdot r_{n-k} r_{n-l} r_{n-m}
\end{aligned}
\end{equation}
where $h_{k}^{(1)}$, $h_{k, l}^{(2)}$ and  $h_{k, l, m}^{(3)}$ are the tap coefficients of $1^{\text{st}}$-, $2^{\text{nd}}$- and $3^{\text{rd}}$-order tap coefficients in $3^{\text{rd}}$-order VNLE, for $i = 1,2,3$, respectively. $K_i$ is the tap number of the $i^{\text{th}}$-order tap coefficients in $3^{\text{rd}}$-order VNLE. Generally, $K_1 \geqslant K_2$ and $K_1 \geqslant K_3$. As a simplified version of $3^{\text{rd}}$-order VNLE, $3^{\text{rd}}$-order PNLE has been proposed to compensate the linear and nonlinear distortions in order to make a trade-off between computational complexity and performance. The output of the $3^{\text{rd}}$-order PNLE is expressed as
\begin{equation}
p_n= \sum_{k=0}^{K_1-1} h_{k}^{(1)}\cdot r_{n-k}+ \sum_{k=0}^{K_2-1} h_{k}^{(2)}\cdot r^{2}_{n-k} + \sum_{k=0}^{K_3-1} h_{k}^{(3)} \cdot r^{3}_{n-k}.
\end{equation}
Apparently, FFE is a special case of PNLE with only the $1^{\text{st}}$-order tap coefficients. One reason why PNLE is the first part of the proposed ACMD is that it can eliminate most of the linear and nonlinear distortions based on the channel characteristics. The second reason is that it can be optimized by adaptive moment estimation (Adam) algorithm, which is much less sensitive to the step sizes \cite{wang2020bgd-based,wang2019optimization,zhou2019adaptive}. The robustness to the step sizes of Adam algorithm is of great importance for the adaptive equalizers, especially when several different step sizes are necessary for the polynomial coefficients with different terms to obtain fast and stable convergence.
\subsubsection{DFE}
In essence, PNLE including FFE as well, is a kind of finite impulse response (FIR) filters. FIR filter is also known as a moving-average (MA) filter, which have only the zeros but no other pole in the Z-plane, except the pole at $z=0$ \cite{ingle2016digital}. However, power-fading effect caused by CD leads to spectral nulls in spectrum of the received signal. The main reason why PNLE including FFE can not completely compensate the power-fading effect induced by CD is that PNLE has no pole to cancel with the spectral zeros introduced by CD. However, infinite impulse response (IIR) filter also known as an autoregressive (AR) filter, has the poles. As a kind of IIR filter combining with a decision-device to achieve the bounded-input bounded-output (BIBO) stability, DFE is able to compensate the distortions caused by spectral nulls more effectively. The best choice of using DFE is the combination of MA filter and AR filter corresponding to an AR-MA filter with both zeros and poles, where MA filter performs precursor equalization and AR filter performs short postcursor equalization. As the part 2 of ACMD, the output of AR-MA type DFE followed PNLE is expressed as
\begin{equation}
q_n = \sum_{k=0}^{F_1-1} f_{1,k} \cdot p_{n-k} + \sum_{l=0}^{F_2-1} f_{2,l} \cdot \hat q_{n-1-l}
\end{equation}
where $f_{1,k}$ is the tap coefficients of $F_1$-tap MA filter and $f_{2,k}$ is the tap coefficients of $F_2$-tap AR filter of DFE. $\hat q_{n}$ denotes decision of  $q_{n}$. Tap coefficients of DFE are trained by using least mean square (LMS) algorithm. After the training process, tap coefficients are tracked by the decision directed-LMS algorithm. One benefit of the use of DFE after PNLE is that compensation for most linear and nonlinear distortions by PNLE makes DFE less prone to error propagation problem.

\subsubsection{MLSE}
Compared to the adaptive equalizers, MLSE has been proved to be  the optimal signal detection for removing ISI distortions in the system with additive white Gaussian noise (AWGN) channel. Although the power-fading effect caused by linear and nonlinear distortions and CD is compensated by the use of PNLE \& DFE, the equalization methods would also inevitably amplify the noise, resulting in colored output noise. The desired expression for the signal-to-noise ratio is \cite{proakis2001digital}
\begin{equation}
SNR=\left[\frac{T^{2} N_{0}}{2 \pi} \int_{-\pi / T}^{\pi / T} \frac{d \omega}{\sum_{n=-\infty}^{\infty}|H(\omega+2 \pi n / T)|^{2}}\right]^{-1}
\end{equation}
where $N_0/2$ is power spectral density (PSD) of the noise, $T$ is the signal duration and $|\omega| \leq \frac{\pi}{T}$. The equalizers attempt to compensate for these distortions by introducing an infinite gain at that frequency at the expense of enhancing the additive noise. Therefore, the performance of the adaptive equalizers is poor whenever the folded spectral characteristic possesses nulls or takes on small values. The output noise would show a spectral profile similar to the joint frequency response of PNLE and DFE. Therefore, the enhanced noise and the residual impairments would still lead to the poor transmission performance. Based on the noise characteristics, a $(P + 1)$-tap noise-whitening PF is adopted to whiten the noise adaptively before implementing optimal signal detection using MLSE. The output of the $(P + 1)$-taps FIR PF can be expressed as
\begin{equation}
v_{n}=\sum_{i=0}^{P} \omega_{i} \cdot q_{n-i}
\end{equation}
where $\omega_{i}$ is the tap coefficients of PF. To design the noise-whitening PF, the Yule-Walker equations for the AR coefficients extraction is adopted \cite{liu2014detection}. AR coefficients extraction using the autocorrelation function of noise provides the way for estimating tap coefficients, which can be calculated by
\begin{equation}
\left[\begin{array}{c}
{\omega_{0}} \\
{\omega_{1}} \\
{\vdots} \\
{\omega_{P}}
\end{array}\right]=\left[\begin{array}{cccc}
{R_0} & {R_1} & {\dots} & {R_P} \\
{R_1} & {R_0} & {\cdots} & {R_{P-1}} \\
{\vdots} & {\vdots} & {\ddots} & {\vdots} \\
{R_P} & {R_{P-1}} & {\cdots} & {R_0}
\end{array}\right]^{-1} \times \left[\begin{array}{c}
{R_0} \\
{0} \\
{\vdots} \\
{0}
\end{array}\right]
\end{equation}
where $R$ is the autocorrelation of noise $z$ . The noise $z_n$ can be estimated as
\begin{equation}
z_{n}=v_{n}-\hat{v}_{n}
\end{equation}
where $\hat{v}_n$ is the decision of $v_n$. It worth noting that the estimate of tap coefficients of the noise-whitening PF can be implemented without training sequence. However, the noise-whitening PF also unavoidably introduces a known ISI. Since after the processing of PNLE \& DFE, there is almost no interference other than noise, the major ISI after the processing of PF is that introduced by the PF. Therefore, the tap coefficients $\omega$ of the PF can be provided for MLSE as the channel information without an extra channel estimator. For OOK modulation, a $2^P$-state MLSE is adopted to eliminate the ISI where $P$ depicts the memory length of MLSE. In a general way, for the M-PAM signal, after filtered by the noise-whitening PF with $(P + 1)$ taps, a $M^P$-state MLSE is used to recover the signal. Under the assumption that the noise after the PF is AWGN, it's equivalent to the minimization of the Euclidean distance between the PF output and transmitted sequence after the PF. In order to detect the most likely transmitted bit sequence from the raw data, the Euclidean distance $D(\boldsymbol{v}, \boldsymbol{s})$ can be expressed as
\begin{equation}
D(\boldsymbol{v}, \boldsymbol{s}) =\sum_{k}[v_k-\sum_{i = 0}^{P} \omega_{i} \cdot s_{n-i}]^{2}.
\end{equation}
The bottleneck of MLSE lies in its exponential increase in computational complexity with increasing memory length $P$. In fact, the high computational complexity of MLSE limits the tap number of the PF. The reason for using MLSE as the last part of the ACMD is that the memory length required for MLSE can be shortened by the PNLE \& DFE and the colored noise can be whitened by PF before implementing optimal signal detection using MLSE.
\begin{figure}
	\centering
	\includegraphics[width = 3IN]{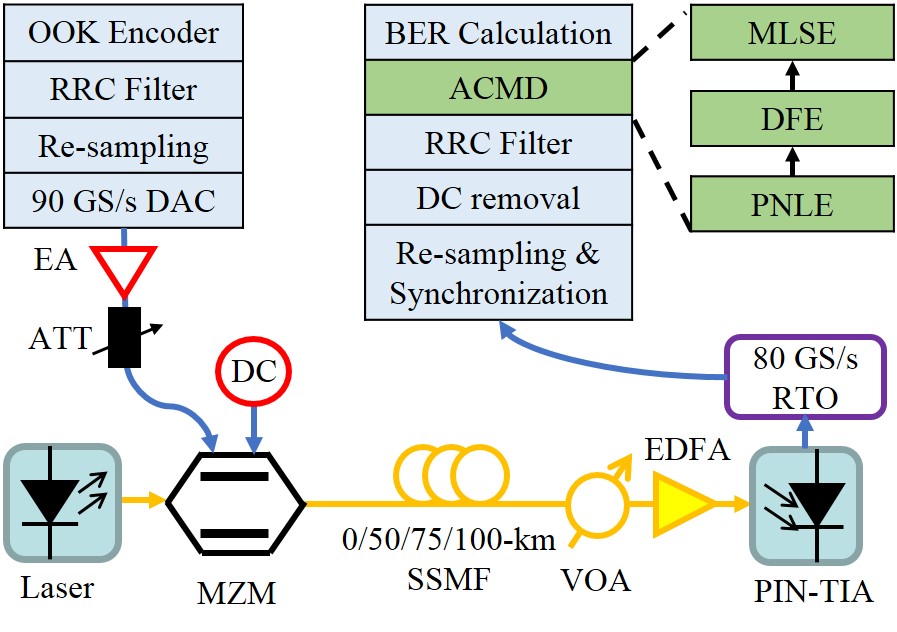}
	\caption{Experimental setup of C-band 64-Gbit/s IM/DD optical OOK transmission over dispersion-uncompensated links using ACMD.}
	\label{EX}
\end{figure}
\begin{figure}[!t]
	\centering
	\includegraphics[width = 3IN]{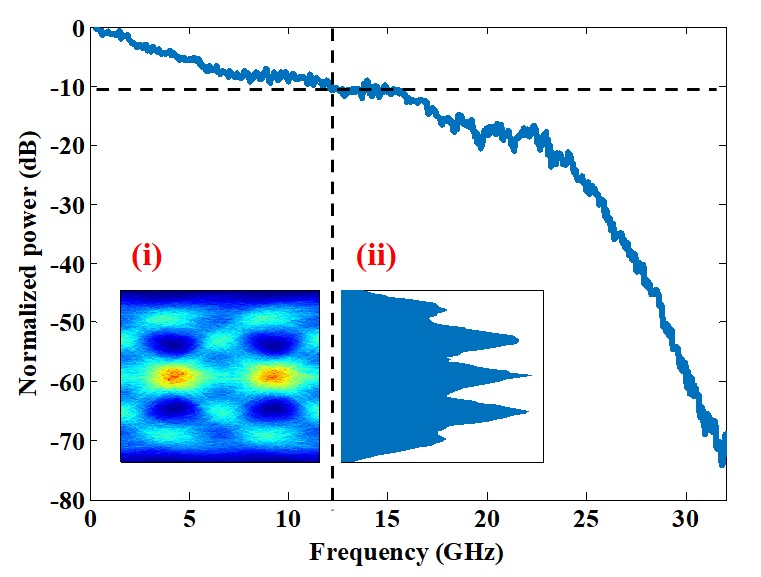}
	\caption{Frequency response of the 64-Gbit/s IM/DD optical OOK OBTB transmission system. Inset (i) is the eye diagram of the received 64-Gbit/s OOK signal over BTB transmission and Inset (ii) is the PDF of the signal.}
	\label{BTB}
\end{figure}

\section{Experimental Setups}
\label{ESetups}
\label{Experimental results}
\begin{figure*}
	\centering
	\includegraphics[width = 7in]{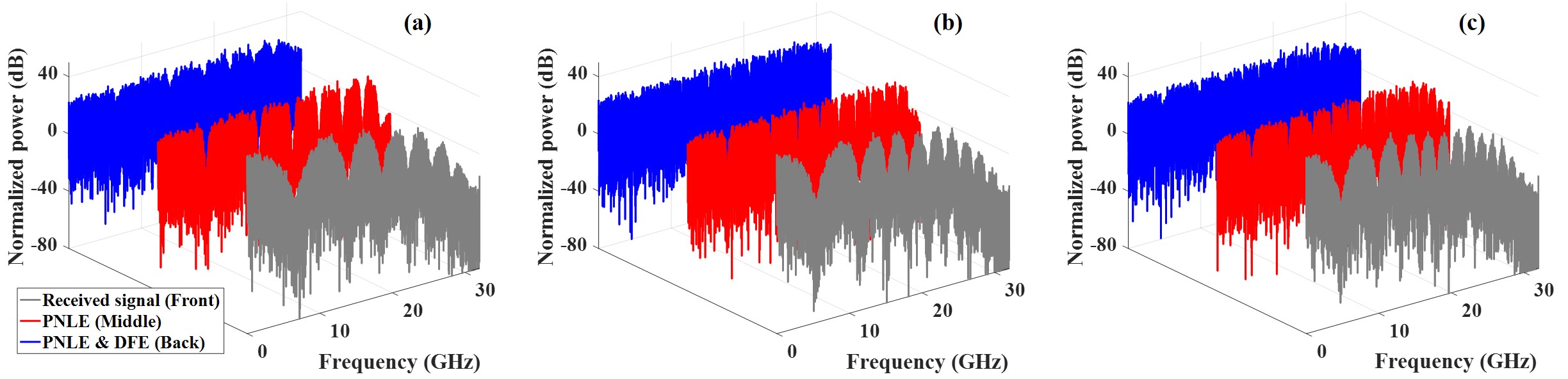}
	\caption{Eletrical frequency spectrum of the received 64-Gbit/s IM/DD OOK signal (in front gray line), PNLE output (in middle red line) and PNLE \& DFE output (in back blue line) over (a) 50-km, (b) 75-km and (c) 100-km dispersion-uncompensated links, respectively.}
	\label{SP}
\end{figure*}

The schematic of the experimental setups of C-band  64-Gbit/s  IM/DD optical OOK transmission over dispersion-uncompensated links is shown in Fig. \ref{EX}. The digital PAM2 (i.e., -1 or 1) frame was generated by off-line processing using MATLAB. One digital PAM2 frame contains 82240 PAM2 symbols. The generated digital PAM2 signals are added a DC component to be unipolar OOK signals. The OOK frame is filtered by the a root-raised-cosine (RRC) filter with the roll-off factor of 0.25 to realize Nyquist pulse shaping and simultaneously reduce ISI. Then the frame is uploaded into digital-to-analog converter (DAC) with 8-bit resolution, 90-GSa/s maximum sampling rate and 16-GHz 3-dB bandwidth. When the sample rate of digital frame was re-sampled to 64-GSa/s, the link rate of generated electrical PAM2 signal was 64 Gbit/s. The base bandwidth of 64-Gbit/s OOK signal is 32 GHz. The net rate was approximately 56.18 Gbit/s (64 Gbit/s $\times$ 77240 / 82240 / (1+7\%) $\approx$ 56.18 Gbit/s) and 55.15 Gbit/s (64 Gbit/s $\times$ 77240 / 82240 / (1+9\%) $\approx$ 55.15 Gbit/s) when the length of training symbols was set to 5000 and FEC with 7\% and 9\% overhead were used, respectively \cite{rafique2014fec}. After being amplified by an electrical amplifier (Centellax OA4SMM4) followed by a 3-dB attenuator (ATT), a 40-Gbps Mach-Zehnder modulator (MZM) @ single drive mode (Fujitsu FTM7937EZ) was used to modulate the amplified PAM2 signal on a continuous wave optical carrier at 1550.116 nm for generating optical OOK signal. A 2V DC bias is applied on the modulator. The generated optical OOK signal was fed into the 0/50/75/100-km SSMF, respectively. The launch optical power was set to 0 dBm for optical back-to-back (OBTB), 7 dBm (i.e., the maximum launch power of the device) for 50-km,75-km and 100-km transmission scenarios, respectively. The link loss was approximately 0.2 dB/km.

At the receiver, a variable optical attenuator (VOA) was used to adjust the received optical power (ROP). Then an EDFA is employed to amplify the signal to ensure that input power of O/E device is enough. The output of EDFA is about -4 dBm. The optical OOK signal was converted into an electrical signal by a 31-GHz P-type-intrinsic-N-type diode with trans impedance amplifier (PIN-TIA) (Finisar MPRV1331A). The electrical signal was fed into a 80-GSa/s real-time oscilloscope (RTO) with 3-dB bandwidth of 36 GHz to implement A/D conversion. The digital OOK signal was decoded by off-line processing, including re-sampling, synchronization, DC removal, RRC matching filter, the proposed ACMD, and bit error rate (BER) calculation.

\section{Experimental Results And Discussion}
\label{ER}
Figure \ref{BTB} shows the frequency response of the 64-Gbit/s IM/DD optical OOK OBTB transmission system. 10-dB bandwidth of the system is about 12 GHz. The channel response fades rapidly when the frequency is beyond 25 GHz. There is no CD induced spectral nulls. Inset (i) is the eye diagram of the received 64-Gbit/s OOK signal and Inset (ii) is the probability distribution function (PDF) of the received signal. Due to the linear and nonlinear distortions, the OBTB system suffers from ISI. The signal amplitude is roughly distributed between -2 and 2. Fortunately, the BER performance can achieve to be below 7\% HD-FEC limit only with the PNLE but no DFE. If a two-tap PF and MLSE with one memory length is used (i.e., $P = 1$), it can achieve to be error-free at ROP of -9 dBm.

Fig. \ref{SP} shows the eletrical frequency spectrum of the received 64-Gbit/s IM/DD OOK signal (in front gray line), PNLE output (in middle red line) and PNLE \& DFE output (in back blue line) over (a) 50-km, (b) 75-km and (c) 100-km dispersion-uncompensated links, respectively.  After the dispersion-uncompensated links transmission, the received signals suffer from quite severe power-fading effect caused by linear and nonlinear distortions and CD. After the processing of PNLE, most of the linear and nonlinear distortions are compensated and the big frequency notches caused by CD are narrowed. However, PNLE has no spectral pole to cancel with the spectral zeros, as analyzed in the Section \ref{ACMDPrinciple}. As a result, the CD-induced spectral nulls still exist. As shown in Fig. \ref{SP}, there are 7, 10 and 14 spectral nulls in the eletrical frequency spectrum of the received signal for 50-km, 75-km and 100-km transmission scenarios, respectively. Fortunately, the CD-induced spectral nulls are compensated by DFE.  After the processing of PNLE \& DFE, the power fading has been greatly compensated. For 50-km transmission scenario, tap numbers $(K_1, K_2, K_3)$ of PNLE are $(111, 71, 41)$ and tap numbers $(F_1, F_1)$ of DFE are $(71, 31)$. For 75-km transmission scenario, $(K_1, K_2, K_3) = (261, 81, 41)$ and $(F_1, F_1) = (71, 41)$, while for 100-km transmission scenario, $(K_1, K_2, K_3) = (291, 81, 41)$ and $(F_1, F_1) = (71, 61)$. Although there is still some weak power-fading effect, in order to make a trade-off between the system performance and computational complexity, the tap numbers of the equalizers do not increase.

\begin{figure}
	\centering
	\includegraphics[width = 3in]{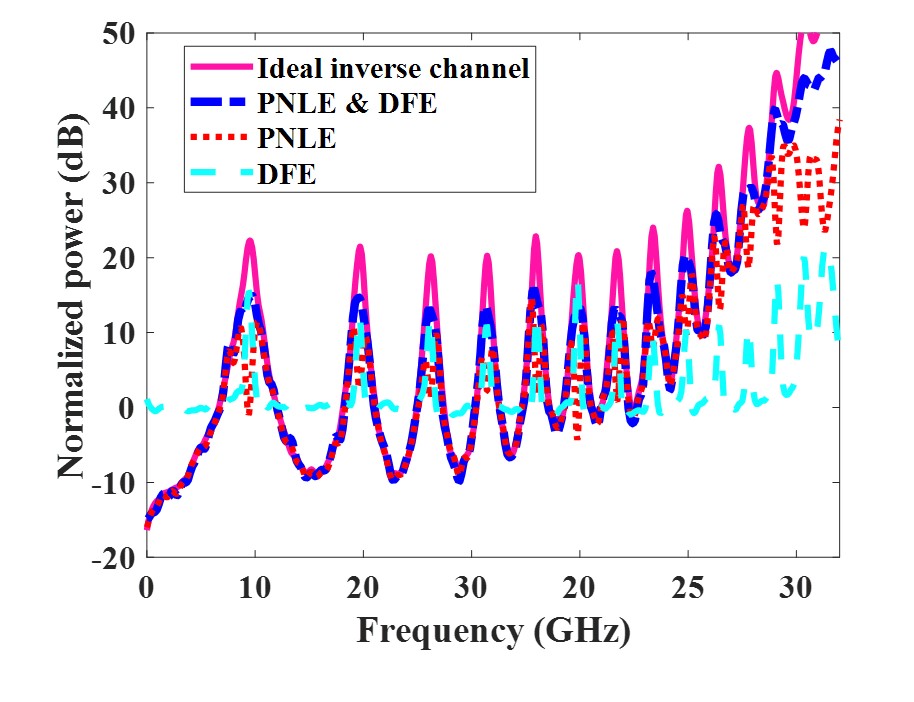}
	\caption{The frequency responses of the ideal inverse channel (magenta solid line),  PNLE \& DFE (blue dotted line), PNLE (red dotted line) and DFE (mint green dotted line) for 64-Gbit/s OOK signal.}
	\label{channel}
\end{figure}

To further understand how PNLE and DFE work, 100-km transmission scenario at ROP of -14 dBm is used to illuminate. Fig. \ref{channel} depicts the frequency responses of the ideal inverse channel (magenta solid line), PNLE (red dotted line), DFE (mint green dotted line) and PNLE \& DFE (blue dotted line) for 64-Gbit/s OOK signal after 100-km transmission scenario. As shown in Fig. \ref{channel}, there are large gains in the frequency response of PNLE near the notches of the received signal spectrum. However, there is no spectral pole in the frequency response of PNLE, as a result of which, there is no infinite gain to compensate system’s frequency response in the locations of CD-induced spectral nulls. Therefore, although the frequency response of PNLE is roughly similar to the ideal inverse channel response, the gain close to the spectral nulls decreases dramatically. Fortunately, there is large gain in the frequency response of DFE to increase the gain of PNLE close to the spectral nulls. The joint frequency response of PNLE \& DFE is quite similar to that of the ideal inverse channel. For the BIBO stability, the feedback signal of DFE is after decision. Therefore, the gain of PNLE \& DFE at the peaks is not very sufficient.

\begin{figure}
	\centering
	\includegraphics[width = 3in]{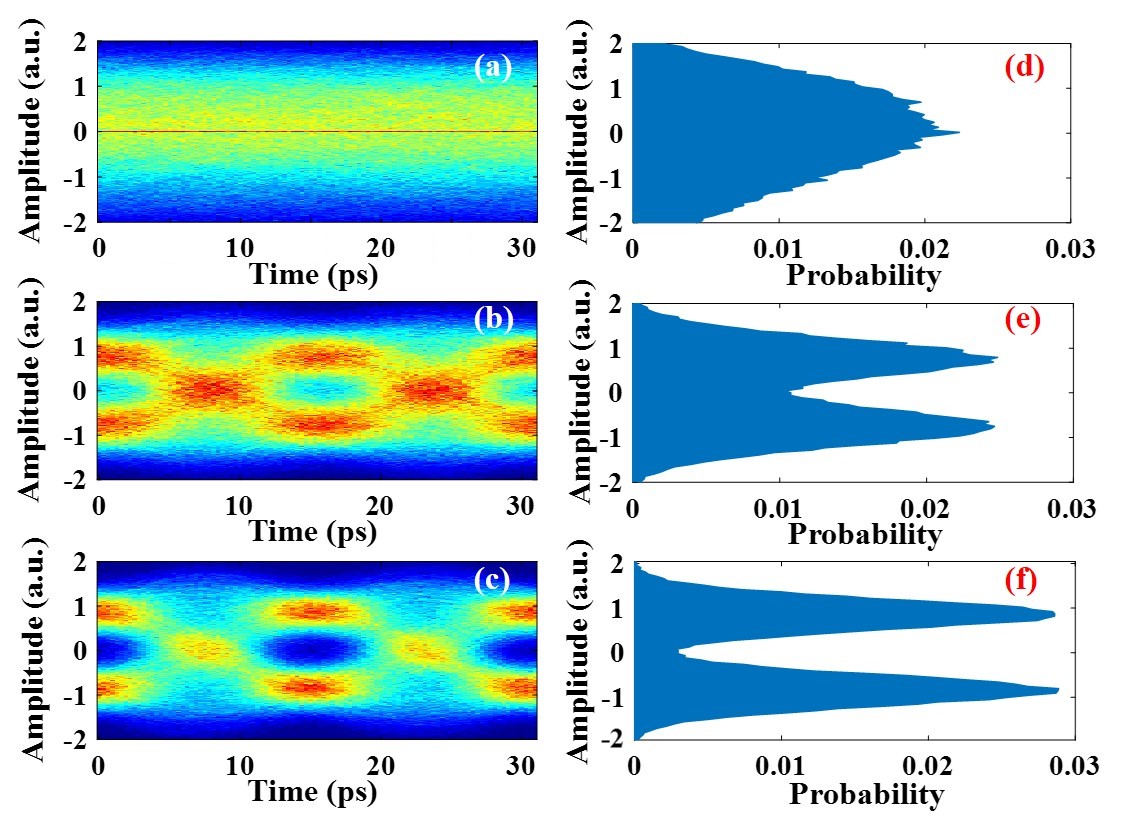}
	\caption{Eye diagram of (a) the received 64-Gbit/s IM/DD OOK signal, (b) PNLE output and (c) PNLE \& DFE output and the signal amplitude PDF of (d) received signal, (e) PNLE output and (f) PNLE \& DFE output.}
	\label{EYE}
\end{figure}

Figure \ref{EYE} show the eye diagram of (a) the received 64-Gbit/s IM/DD OOK signal, (b) PNLE output and (c) PNLE \& DFE output and the signal amplitude PDF of (d) received signal, (e) PNLE output and (f) PNLE \& DFE output. Compared to the eye diagram and the PDF of the received signal in OBTB transmission scenario shown in Inset(i) and (ii) of Fig. \ref{BTB}, the received signal suffered from more severe distortions due to CD in 100-km transmission scenario. The eye of the received signal in 100-km transmission scenario is almost close and the amplitude mostly distributes around 0. After the processing of PNLE, most linear and nonlinear distortions are eliminated. The eye diagram of PNLE output is open and the signal most distributes around -1 and 1. However, due to the power-fading effect caused by spectral nulls, there is still residual ISI. A significant portion of the signal distributed around 0. After eliminating the spectral nulls by DFE, the eye became clearer and most the residual ISI is cancelled. After the processing of PNLE \& DFE, the signal distributed mostly in -1 and 1.

\begin{figure}
	\centering
	\includegraphics[width = 3.1in]{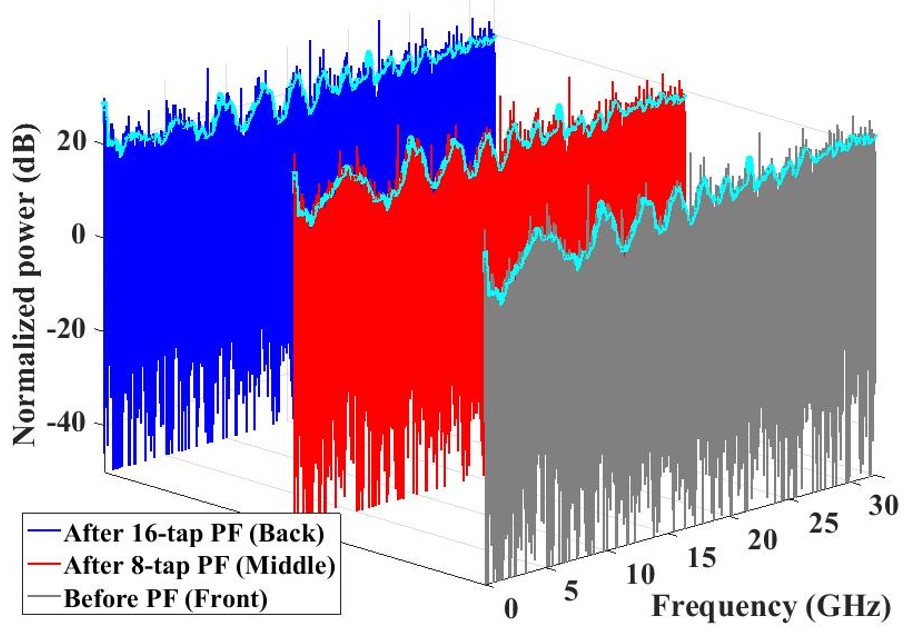}
	\caption{PSD of estimated noise before PF (in front gray line), after 8-tap PF (in middle red line) and after 16-tap PF (in back blue line) for 100-km transmission scenario at ROP of -14 dBm.}
	\label{NOISE}
\end{figure}

There are two solutions to further improve the system performance. The main factor that limits the system BER performance for 100-km transmission scenario is OSNR. Therefore one solution is to increase the launch power. However, due to the power limitation of the devices used in the experiment, we can not increase the launch power. Another way to further improve system performance is to increase the taps number of the noise-whitening PF. Fig. \ref{NOISE} shows the PSD of estimated noise before PF (in front gray line), after a 8-tap PF (in middle red line) and after a 16-tap PF (in back blue line) for 100-km transmission scenario at ROP of -14 dBm. The estimated noise before the PF shows a spectral profile similar to the joint frequency response of PNLE \& DFE. Therefore, a noise-whitening PF is adopted to suppress and whiten the noise, paving the way for for implementing optimal signal detection using MLSE to further improve the system performance at the last part of ACMD. After the processing of a 8-tap PF, the high frequency noise is suppressed, while that in low frequency is slightly amplified compared to the PSD of estimated noise before the PF. The noise in the low frequency is further amplified and the components in the first peak are suppressed if a 16-taps PF is used. Compared with the red line envelope of the PSD of estimated noise after a 8-tap PF in Fig. \ref{NOISE}, the blue line envelope of the PSD of estimated noise after a 16-tap PF is more flatten. Therefore, the noise-whitening effect of 16-tap PF is more obvious than that of the 8-tap PF.

\begin{figure}
	\centering
	\includegraphics[width = 3IN]{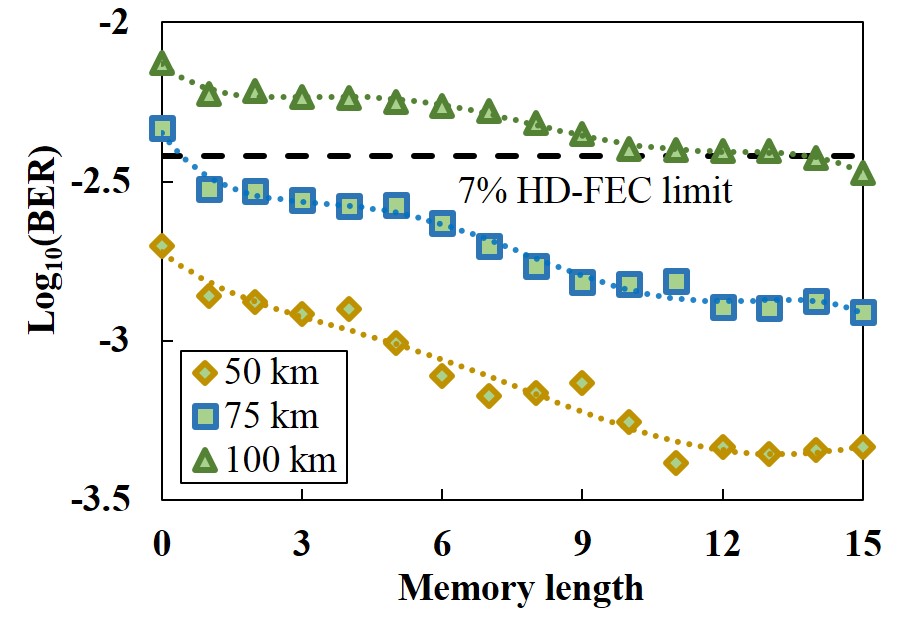}
	\caption{BER performance versus memory length of MLSE for 64-Gbit/s IM/DD optical OOK system after 50-km, 75-km, and 100-km SSMF transmission, respectively.}
	\label{MEMLEN}
\end{figure}

\begin{figure*}
	\centering
	\includegraphics[width = 7in]{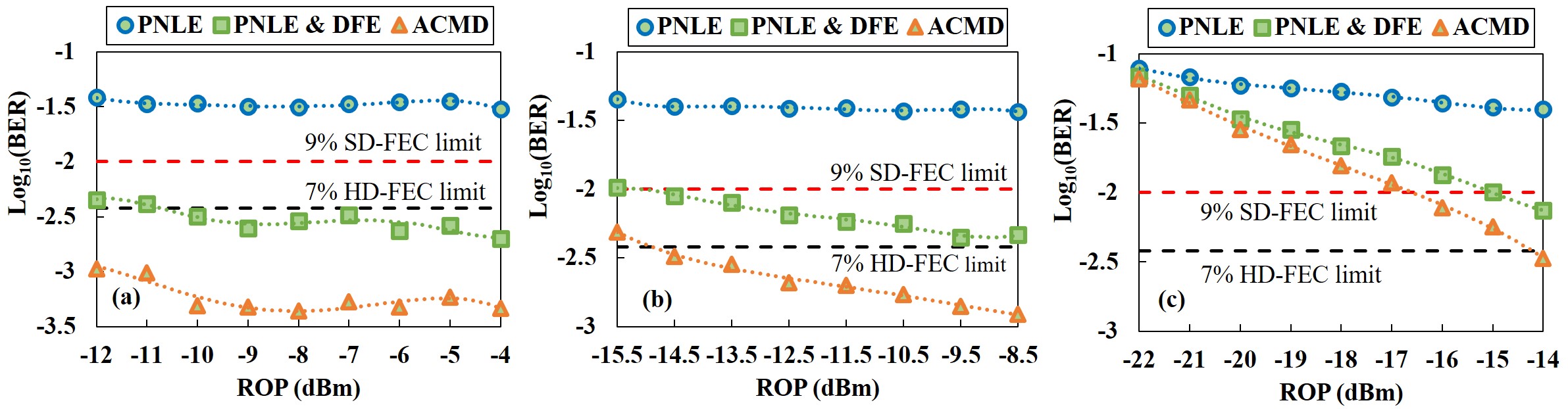}
	\caption{BER performance for 64-Gbit/s IM/DD optical OOK after the processing of PNLE (blue circle), PNLE \& DFE (green square) and ACMD (orange triangle) versus ROP for (a) 50-km, (b) 75-km and (c) 100-km transmission scenarios, respectively.}
	\label{BvP}
\end{figure*}

Figure \ref{MEMLEN} shows the BER performance versus memory length $P$ of MLSE for 64-Gbit/s IM/DD optical OOK system after 50-km, 75-km, and 100-km SSMF transmission, respectively. The ROPs are -4 dBm, -8.5 dBm and -14 dBm for 50-km, 75-km, and 100-km transmission scenarios, respectively. It means that the noise-whitening PF and MLSE are not used when memory length $P$ is equal to zero. As the memory length of MLSE increases, more taps can be used for the PF. Therefore, the system BER performance improves with the improvement of noise-whitening effect. When the memory length $P$ of MLSE is equal to or greater than 14, the system BER performance achieves to be below the 7\% HD-FEC limit for 100-km transmission scenario. BER performance is near $10^{-3}$ and $10^{-3.5}$ for 75-km and 50-km transmission scenarios when the the memory length $P$ of MLSE is equal to or greater than 12 and 11, respectively.

Figure \ref{BvP} shows the BER performance for 64-Gbit/s IM/DD optical OOK after the processing of PNLE (blue circle), PNLE \& DFE (green square) and ACMD (orange triangle) versus ROPs for (a) 50-km, (b) 75-km and (c) 100-km transmission scenarios, respectively. For 50-km and 75-km transmission scenarios with enough ROP, BER of signal after the PNLE can decrease to $\sim 10^{-1.5}$ from $\sim 10^{-0.31}$ (i.e., $\sim 0.5$). For 100-km transmission scenario, BER performance of signal after the PNLE improves as ROP increases and reaches $\sim 10^{-1.4}$ when ROP is -14 dBm. After the processing of PNLE \& DFE, as expected, BER falls sharply. BER performance of signal after the PNLE \& DFE can achieve to be below 7\% HD-FEC limit for 50-km transmission scenario and 9\% SD-FEC limit for 75-km and 100-km transmission scenarios. Obviously, as the first-two parts of the proposed ACMD, PNLE \& DFE can eliminate most linear and nonlinear distortions adaptively based on the channel characteristics. The tap numbers of PF are all set to 16 for 50-km, 75-km and 100-km transmission scenarios. Based on the noise characteristics, the PF can adaptively whiten the noise for implementing optimal signal detection using MLSE. After the processing of ACMD, BER can achieve to be near $10^{-3.5}$ for 50-km transmission scenario and below 7\% HD-FEC limit for 75-km and 100-km transmission scenarios.

\section{Conclusion}
\label{Conclusion}
In this paper, we propose ACMD for C-band 64-Gbit/s IM/DD optical OOK system over 100-km dispersion-uncompensated link. The proposed ACMD can adaptively compensate most of the link distortions based on channel and noise characteristics. ACMD includes a PNLE, a DFE and MLSE. Based on the channel characteristics, PNLE eliminates the linear and nonlinear distortions, while the followed DFE compensates the spectral nulls caused by CD. Finally, based on the noise characteristics, a PF can whiten the noise for implementing optimal signal detection using MLSE. To the best of our knowledge, for the first time, we present a record C-band 64-Gbit/s IM/DD optical OOK system over a 100 km dispersion-uncompensated link with the BER performance below 7\% HD-FEC limit using only ACMD at the receiver side. If the bandwidths of DAC and oscilloscope are increased, the data rate is expected to increase further. In conclusion, ACMD-based C-band 64-Gbit/s optical OOK system shows great potential for future optical interconnects.


%




\ifCLASSOPTIONcaptionsoff
  \newpage
\fi



\bibliographystyle{IEEEtran}
\bibliography{sample}
\end{document}